\documentclass[epj]{svjour}
\usepackage{graphics, amsmath, amssymb, bm}
\begin{document}
\title{Band Gap Modulation of Graphene on SiC}

\author{Stefan Kolev\inst{1} \and Victor Atanasov\inst{2} \and Hristiyan Aleksandrov\inst{3} \and Teodor Milenov\inst{1}% etc
% \thanks is optional - remove next line if not needed
%\thanks{\emph{Present address:} Insert the address here if needed}%
}                     % Do not remove
\offprints{}          % Insert a name or remove this line
\institute{Institute of Electronics, Bulgarian Academy of Sciences, 72 Tzarigradsko Chausee Blvd., 1784 Sofia, Bulgaria \and Faculty of Physics, Sofia University, 5 blvd J. Bourchier, Sofia 1164, Bulgaria \and
Faculty of Chemistry and Pharmacy, Sofia University, 1 J. Bourchier Blvd., Sofia 1164, Bulgaria
}
%
%\date{Received: date / Revised version: date}
% The correct dates will be entered by Springer
%
\abstract{
A recipe on how to engineer a band gap in the energy spectrum for the carriers in graphene is conveyed. It is supported by a series of numerical simulations inspired by an analytical result based on the opening of a band gap in periodically corrugated graphene, e.g. the buffer layer grown on SiC at high temperatures. \\
{\bf Key Words:} {\it graphene, band gap, ab initio simulations, semiconductors, buffer layer, SiC}
\PACS{
      {73.22.Pr}{}   \and
      {73.61.Wp}{} \and {71.20.-b }{}
     } % end of PACS codes
} %end of abstract
\maketitle
\section{Introduction}
\label{intro}
Graphene has been arround for more than 15 years now \cite{[1]}. Despite its aparent excellent electronic characteristics \cite{[2]} it remains a material largely unsuitable for electronics aplications, mainly due to its semi-metallic character, i.e. it has a vanishing band gap in its electronic structure. As a result, if one attempts to construct a field effect transistor (FET), one would end up with a device which has no ON/OFF state. Graphene is unable to replace silicon, even though it seems the be the next logical step mainly due to the ballistic transport and aweinspiring carrier mobility of 200 000 cm$^2$ V$^{-1}$ s$^{-1}$ \cite{[3]}. Yet, the prized aspect of graphene – carrier mobility (high operating frequencies of future graphene based FETs) is completely shadowed by its lack of a band gap.

The situation is made even more dramatic by the absence of a proven technology for its wafer-scale production with sufficient quality \cite{[4]}. Not only that but the ideal technology should be able to place it ontop of a sufficiently thin insulating layer. Yet, the main candidate for large scale wafer-sized production, that is chemical vapour deposition (CVD) and its variations and upgrades, is dealing poorly with making it ontop of an insulating support \cite{[5]}. The vicious circle closes in, when one is reminded of the other possible wafer-scale technology: graphitization of SiC at high temperatures \cite{[6],[7]}. Here, the devil hides in the following detail, the SiC that remains as a support to the graphene layers is either semiconducting (2.36 eV band gap for 3C($\beta$)) or insulator (3.23 eV band gap for 4H; 3.05 eV band gap for 6H($\alpha$) ). It is way too thick to be any good for inducing the field effect necessary to swich ON/OFF the FET’s channel.

However, it is worth exploring any vanue that might lead to opening of a band gap in graphene, thus turning it into a semiconductor. Yet, we should be aware that this transformation might come at a cost, which is to be paid by sacrificing carrier mobility. How big this loss of carrier mobility might be, remains an open question. 

A couple of years ago, a breakthrough in graphene electronics research, namely that the first graphene layer grown on the SiC(0001) surface, that is the "buffer" layer, can be prepared as a true semiconductor with a band gap of  $>$0.5 eV was reported \cite{[8]}. This band gap is a result of an improved order in the formation of covalent bonds between the buffer layer and the SiC, caused by higher growth temperature. This buffer graphene, commensurately bonded to the SiC(0001) surface, is an example of a functionalized system, where periodic bonding to either all A or all B sites breaks graphene's chiral symmetry \cite{[9]}. The authors offer an explanation for the gap opening based on finite size effect \cite{[10]}, which certainly does not lead to an applicable model predicting the opening of a band gap with prescribed size. 

\section{Computational methods}
\label{sec:2}
In order to refine and confirm the model described below, we performed {\it ab initio} simulations of the different configurations: pristine graphene; SiC covalent bonded to corrugated or non-corrugated graphene; and SiC covalent bonded to corrugated graphene with additional graphene layers.

{\it Ab Initio} geometry optimizations were performed using the CP2K/Quickstep package \cite{[11],[12]}. The DFT was applied within the generalized gradient approximation
 (GGA), using Perdew-Burke-Ernzerhof (PBE) functional \cite{[13]}. Basis set DZVP-MOLOPT-SR-GTH \cite{[14]}, which is optimized for calculating molecular properties in gas and condensed phase, was applied for all atoms in the studied systems. For reducing computational cost Gaussian and Plane-Wave (GPW) method was used \cite{[15],[16]}. This method uses an atom-centered Gaussian-type basis to describe the wave functions and an auxiliary plane wave basis to describe the electron density. Only the valence electrons are explicitly treated. Their interaction with the remaining ions is described using the pseudopotentials of Goedecker-Teter-Hutter (GTH) \cite{[17],[18]}. The charge density cutoff of the finest grid level is equal to 400 Ry. The number of used multigrids is 5. Dispersion interaction is taken into account via the DFT + D approach, with D3 set \cite{[19]}.

Convergence of the SCF procedure is usually difficult to achieve in systems with small band gap or isoenergetic states as metals and semimetals. This difficulty arises from having to integrate discontinuous functions that drop to zero when the valence band crosses the Fermi energy \cite{[20]}. The difficulty to converge the SCF procedure increases with increasing the complexity of the system.

In order to improve convergence with respect to Brillouin zone sampling in the systems with small band gap, electronic temperature was introduced, using the Fermi-Dirac distribution method \cite{[20]}. 
\begin{equation}
f(E) = \frac{1}{e^{(E-E_f)/kT}+ 1}
\end{equation}
Where $f(E)$ is the probability that an electron will have energy $E,$ $E_f$ is the Fermi energy at $T = 0$K, $k$ is the Boltzmann constant and $T$ is the temperature in K. All calculations were performed at $T = 300$K and 30 additional unoccupied orbitals were added. Smearing of the MO occupation numbers led to partial occupation of orbitals close to the Fermi energy. This method allowed for achieving convergence in all cases.

Trust (maximum) radius of the geometry optimization step was initially set to 5 pm, but in some cases oscillation of geometry and energy occurred. In these cases the trust radius was reduced to 1 pm in order to achieve convergence. 

Radial distribution functions (RDF) are plotted with a step of 1 pm. For all systems, the density of states was calculated with VASP software \cite{[21],[22]}, employing k-point sampling with 7 x 7 x 1 grid. The same level of DFT was used with both the VASP and CP2K calculations. Band gap values are calculated directly from the electronic structure of the systems, as HOMO-LUMO energy difference, without considering peaks broadening.

The systems with non-corrugated and semi-corrugated buffer layer graphene (BuLG) are derived from the system with corrugated BuLG. In the first case, all values of z-axis (graphene C atoms) are made equal to constant, creating a flat system without corrugation. In the second case, all values of z-axis are divided by 2, reducing the corrugation by half as $.5 z_2 - 0.5 z_1 =.5 (z_2 - z_1).$ Si-C(graphene) bonds are kept in the interval 195-210 pm so the electron effect of SiC proximity remains equal for all systems. Geometry of all systems with bilayer graphene, including SiC substrate or not, is optimized from initial AB stacked configuration. Geometry of the system with three graphene layers is optimized from initial ABA stacked configuration. 

Due to the large size of the studied systems, the simulations described below took about 600 000 CPU*hours to be completed (2 GHz CPUs, 32 or 64 in parallel).

\section{Motivation}
\label{sec:3}

One step closer to a model having band gap can be brought by exploring the following geometrically inspired hypothesis: the energy gap is geometric in nature and a result of a periodic corrugation of the buffer graphene due to the strain induced by the covalent bonding with the substrate (normal to the sheet’s plane). As it is well known, the electronic system of graphene is relativistic in nature and therefore hardly sensitive to local defects \cite{[23],[24]}. However, periodically spaced sp$^3$ hybridised carbon atoms, covalently bridging the layer to the SiC substrate, force the graphene sheet into a periodic corrugation. This periodic corrugation can be revealed by a HRTEM study of a cross-section of the SiC wafer (as seen in \cite{[25]}), from which the typical values for the period $a \approx 2.4 - 2.6 \AA$ and height $h \approx 0.4 - 0.8 \AA$ can be extracted.

\begin{table}[b]
\centering
\begin{tabular}{|c|c|c|c|c|}
\hline
$E_{gap}$\; {\rm [eV]} & \multicolumn{4}{c|}{$a$ {\rm [$\AA$] }  } \\ \hline
                             & 1.5       & 2.4       & 3         & 4.5      \\ \hline
0.26                         &  {\bf 0.3}       & {\bf 0.6}       & {\bf 0.8}       & {\bf 1.5}      \\ \hline
0.5                          & {\bf 0.4}       & {\bf 0.8}       & {\bf 1.2}       & {\bf 2.2}      \\ \hline
\end{tabular}
\caption{The estimated values of the necessary height $h$ of the corrugation in Angstrom, to produce the corresponding energy gap.}
\label{table1}
\end{table}

The proof-of-concept calculation relies on two results: i.) the electronic properties of the periodically corrugated in one dimension graphene are governed by the Mathieu equation \cite{[26]}:
\begin{equation}
\left[ d^2/dz^2 + \alpha -2 q \cos(2z)  \right] \phi(z)=0,
\end{equation}
where $z=s/a$ is a dimensionless parameter with the meaning of length, since $s$ is the arc-length along the corrugation and $a$ is the period of the corrugation. $\alpha=a k_s$ is again dimensionless and here $k_s$ is the wavenumber along the arc-length. $q=-h^2 /(8a^2 ),$ where $h$ is the height of the corrugation; ii.) the $n-$th gap ($n\in \mathbb{N}$) $\Delta_n$ of the Mathieu equation is given by\cite{[27]}: 
\begin{equation}
\Delta_k \approx 8 (|q|/4)^n * 1/[(n-1)!]^2. 
\end{equation}

Inserting the relevant quantities from i.) into ii.)  for $n=1$ we obtain:
\begin{equation}
E_{gap} = \frac{\hbar v_F}{4a} \left(\frac{h}{a}\right)^2   \approx \left(\frac{h}{a}\right)^2 \frac{10}{a}\;  {\rm [eV]},
\end{equation}
where the formula on the right works with $a$ in Angstrom. Note $\hbar$ is Planck's constant and $v_F \approx 10^6$[m/s] is the Fermi velocity in graphene. The necessary height $h$ of the corrugation in Angstrom, to produce the corresponding gap is presented in Table \ref{table1}.

In order to produce the energy gap reported in \cite{[8]}, the buffer layer needs to be corrugated with a period of $2.4 \AA$ and a height of $0.8 \AA$, slightly more than the Bohr radius and well within experimentally set bounds: the distance between SiC-buffer layer is approximately $2.9 \AA$; the distance between 1st graphene layer and the 2nd graphene layer is approximately $3.7 \AA$ \cite{[25]}. Therefore, the strained covalently bonded buffer layer can be corrugated by as much as $0.8 \AA$, literally what is necessary to produce the observed gap.

Having this motivation in mind, we proceed to further elaborate on the idea that a periodically spaced covalent bonding to the graphene sheet leads to corrugation and subsequent opening of a band gap in the energy spectrum for the carriers.

\begin{figure*}[t]
% Use the relevant command for your figure-insertion program
% to insert the figure file.
% For example, with the option graphics use
\resizebox{0.95\textwidth}{!}{%
  \includegraphics{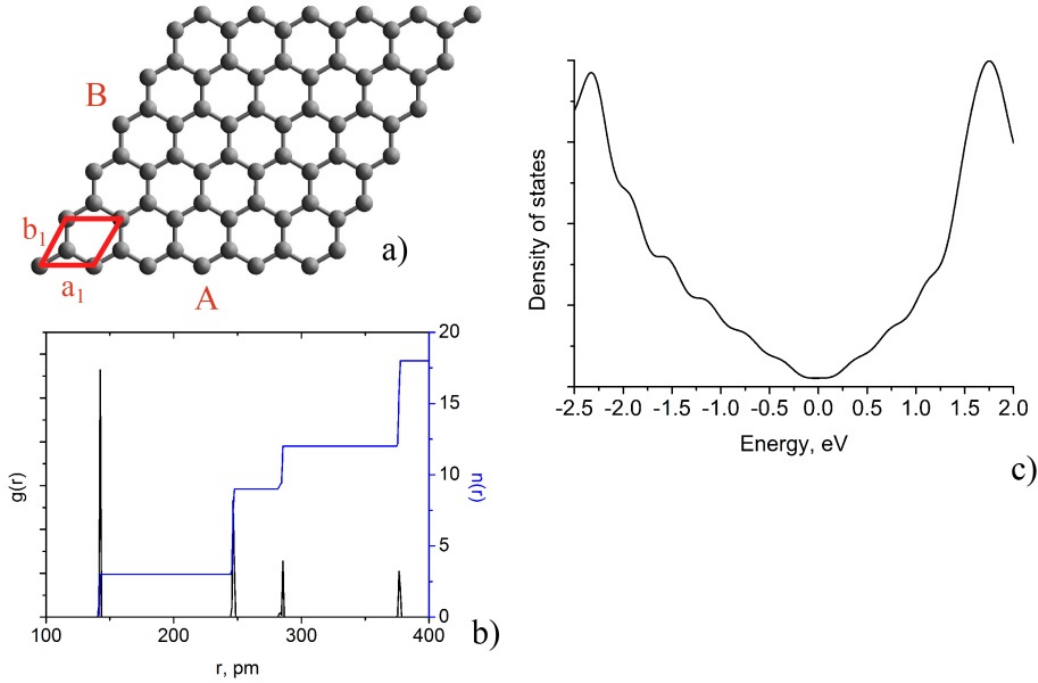}
}
% If not, use
%\vspace{5cm}       % Give the correct figure height in cm
\caption{ a) Periodic system consisting of 72 carbon atoms after the geometry optimization; b) Radial distribution function (RDF) - black trace of one carbon atom versus the whole system and integral (blue trace) equal to the number of carbon atom neighbors; c) The dispersion of the density of states - the valence band is practically overlapped with the conduction band. }
\label{fig1}       % Give a unique label
\end{figure*}

\section{Main hypothesis}
\label{sec:4}

We propose and study a simplified model in order to investigate the opening of a band gap in the graphene buffer layer, i.e. the immediate graphene film on top of the (0001)$_{\rm Si}$ SiC surface. In this model, we take into account the influence of the covalent bonds between the carbon (from graphene) and silicon (from SiC) atoms at the interface, on the physical properties of the graphene film, i.e. the hybridization of the directly bonded carbon atoms shifts from sp$^2$-hybridization (in graphene) to sp$^3$- hybridization (in diamond-type structure). Such a shift leads to electronic and structural effects, since the carbon atoms in sp$^3$ hybridization do not participate along with {\it p} electrons in the global delocalized {\it p} electron density of states for graphene. Thus, we come to the main hypothesis in the paper, namely a pattern of periodically distributed onto the C/SiC interface sp$^3$ hybridized carbon atoms causes a periodic corrugation of the sheet which leads to an opening of a band gap, i.e. semiconducting graphene. A similar model was studied theoretically and experimentally earlier \cite{[28],[29],[30],[31]} and the results point to the graphene layer on top of the SiC surface being regarded as a buffer layer, i.e. it has different properties than that of the following graphene layer/s. The simulations performed by Varchon et al. \cite{[28]} used a slab of eight 4H- SiC cells (3 x 3 R30 surface cell in Wood’s notation) and situated one (buffer) as well as two (one buffer and first graphene) graphene layers (1 x 1 surface cell) on top of the SiC. It should be noted that the 2 x 2 surface cell of graphene almost matches the 3 x 3 R30 surface cell of SiC with an only 8\% mismatch between the lattices and it is established that it has no qualitative effect on a freestanding graphene electronic structure (it can, however, change the Dirac electron velocity \cite{[28]}). Additionally, the authors saturated the dangling bonds by hydrogen atoms (one dangling bond on a 2 x 2 graphene cell) and showed that an energy gap is opened in the buffer layer [28]. However, the relation between sheet corrugation and band gap opening is explored using {\it ab initio} methods, for the first time in the present publication.

\section{Structure model of graphene and {\it ab initio} simulations}
\label{sec:5}

We proceed to the numerical experiment establishing a basic structural model. The model of graphene is based on a hexagonal cell that consists of two carbon atoms and has dimensions (${\bf a_1} = {\bf b_1}$) of 246 pm and angle $\gamma$= 60 deg, see Fig. \ref{fig1}, where the red lines mark the unit cell. The unit cell can be represented in Wood’s notation as 1 x 1 cell in the coordinates of the graphene cell. The supercell used in the {\it ab initio} simulations of graphene sheet contains 72 carbon atoms. After the geometry optimization it has dimensions of {\bf A} = 1480 pm, {\bf B} = 1480 pm and {\bf C} = 1900 pm. The angles between {\bf A},{\bf B} and {\bf C} edges of the supercell are $\alpha$ = 90 deg - between {\bf A} and {\bf C}, $\beta$ = 90 deg - between {\bf B} and {\bf C} and $\gamma$ = 60 deg - between {\bf A} and {\bf B} (depicted in Fig. \ref{fig1}). The volume along the {\bf C} - edge above the basic graphene layer remains unoccupied and is assumed large enough to eliminate the interaction between the graphene layers in the periodic system. This geometry of the system corresponds to C-C interatomic distance of mainly 142 pm- see panel b in Fig. \ref{fig1}, which was the mean value reached after 1 ps dynamical simulation of a larger graphene sheet \cite{[32]}. The molecular orbitals are calculated from the optimized geometry. All occupied orbitals as well as the first 100 unoccupied are taken into account. The energy gap (difference between the highest occupied orbital (HOMO) and the lowest unoccupied orbital (LUMO) is negligible and equals to $\Delta E$=0.06 eV. This small value is indistinguishable as it is seen in the density of states - the c) panel of the Fig. \ref{fig1}.

\section{Structure model of graphene on (0001)$_{\rm Si}$ plane of 6H-SiC}
\label{sec:6}

\begin{figure*}
% Use the relevant command for your figure-insertion program
% to insert the figure file.
% For example, with the option graphics use
\resizebox{0.95\textwidth}{!}{%
  \includegraphics{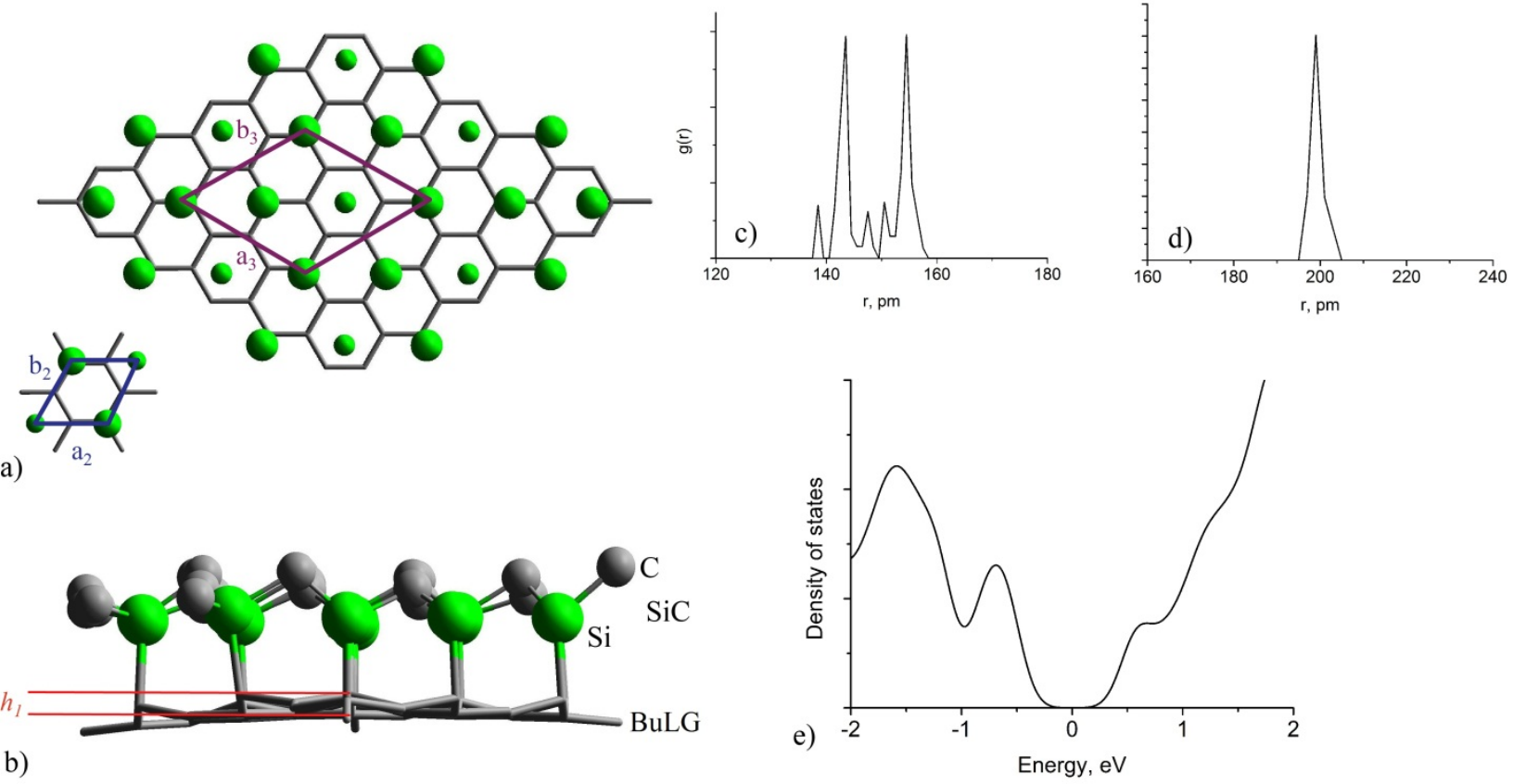}
}
% If not, use
%\vspace{5cm}       % Give the correct figure height in cm
\caption{ a) The superposition of the basic structure models, viewed along the c- axis: the dark grey- sticks model represent the graphene sheet (a 1 x 1 surface cell (${\bf a_1 = b_1}$)), the green spheres model represents the Si- atoms of 3 x 3 R30 surface cell (with basic vectors ${\bf a_2 = b_2}$ remarked with blue lines, the inset in lower right corner) of SiC. The smaller green spheres are expected to have dangling bonds and are excluded from our structure model. Thus the larger green spheres form a new 2 x 2 (in graphene) surface cell with ${\bf a_3 = b_3}$ basic vectors (purple lines); b) The optimized structure SiC/ buffer graphene layer viewed along the direction perpendicular to the bisection of the angle between {\bf A} and {\bf B} axis. BuLG acronym designates the buffer layer graphene and $h_1$ is the height of the corrugation of BuLG ($h_1$ = 35.0 ± 5.0 pm); c) Radial distribution function of graphene carbon atoms, the C-C interatomic distance in sp$^2$-hybridized carbon is 143 pm, while the C-C interatomic distance (sp$^3$-hybridized carbon) is 154 pm; d) Radial distribution function of sp$^3$ hybridized carbon atoms and silicon atoms; e) The dispersion of the density of states.  }
\label{fig2}       % Give a unique label
\end{figure*}

Next, we proceed with the numeric experiment by assuming that the influence of the third and higher order neighbors of the Si-atoms at the interface SiC/ graphene on the density of states of the buffer layer graphene, respectively, could be neglected. Therefore, in our structure model, we use a single atomic layer of C atoms from graphene, their first neighboring Si atoms and the first C atoms from the SiC substrate (Fig. \ref{fig2}a). Hydrogen atoms (not shown in Fig. \ref{fig2}a) are employed to saturate the remaining free bonds of C (from the SiC substrate). The Si layer is built based on the unit cell of SiC \cite{[33]} and has a 3 x 3 R30 surface cell (corresponding to a 2 x 2 graphene surface cell) - the green spheres model and the blue line remark the unit cell in the inset in the right lower corner of Fig. \ref{fig2}a. It is concluded that, along with the natural mismatch of the lattices, graphene has a unit cell, which is 30 deg rotated with respect to the unit cell of the SiC substrate \cite{[29]}. This rotation is taken into account in our model. The Si layer is situated on top of the carbon atoms of graphene and we remove the Si atoms with dangling bonds (smaller green spheres in Fig. \ref{fig2}a) in our further considerations. It should be noted that the valent electrons of the silicon atoms, bonded to carbon atoms from graphene layer, point at angles close to 90 deg away from the surface, i.e. it is expected that the carbon atoms should be sp$^3$ hybridized and situated below the Si- atoms in order to be covalently bonded. The remaining electrons of the Si atoms form Si-C covalent bonds in the SiC substrate.

\begin{figure}[h]
% Use the relevant command for your figure-insertion program
% to insert the figure file.
% For example, with the option graphics use
\resizebox{0.45\textwidth}{!}{%
  \includegraphics{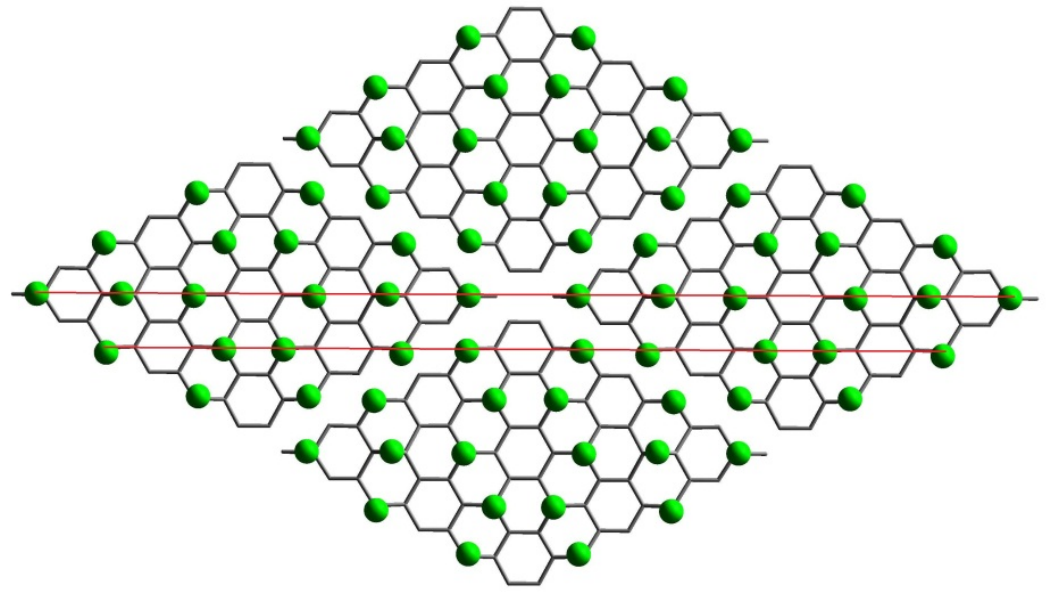}
}
% If not, use
%\vspace{5cm}       % Give the correct figure height in cm
\caption{ Four images of the periodic cell (+x; +y representation). Red lines show the corrugation, caused by covalent bonded Si atoms, running through the periodic system.   }
\label{fig3}       % Give a unique label
\end{figure}

The system consists of a buffer graphene layer (BuLG) that contains 72 carbon atoms, bonded to a SiC layer, Fig. \ref{fig2}b. The system is in singlet ground state with total spin equal to zero. A geometry optimization is performed and the individual positions of the atoms as well as the dimensions of the periodic cell are optimized. Positions of the silicon atoms are fixed during the geometry optimization. The cell has dimensions {\bf A} = 1529 pm; {\bf B} = 1529 pm and {\bf C} = 1716 pm. The angles between {\bf A}, {\bf B} and {\bf C} edges of the periodic cell are as follows: $\alpha$ - between {\bf A} and {\bf C}, $\beta$ - between {\bf B} and {\bf C}, and $\gamma$ - between {\bf A} and {\bf B} with values of 90 deg, 90 deg and 60 deg, respectively. The geometry and symmetry of the cell are chosen, in order to preserve the line of corrugation throughout the whole system of periodic images, Fig \ref{fig3}.

\begin{figure}
% Use the relevant command for your figure-insertion program
% to insert the figure file.
% For example, with the option graphics use
\resizebox{0.4\textwidth}{!}{%
  \includegraphics{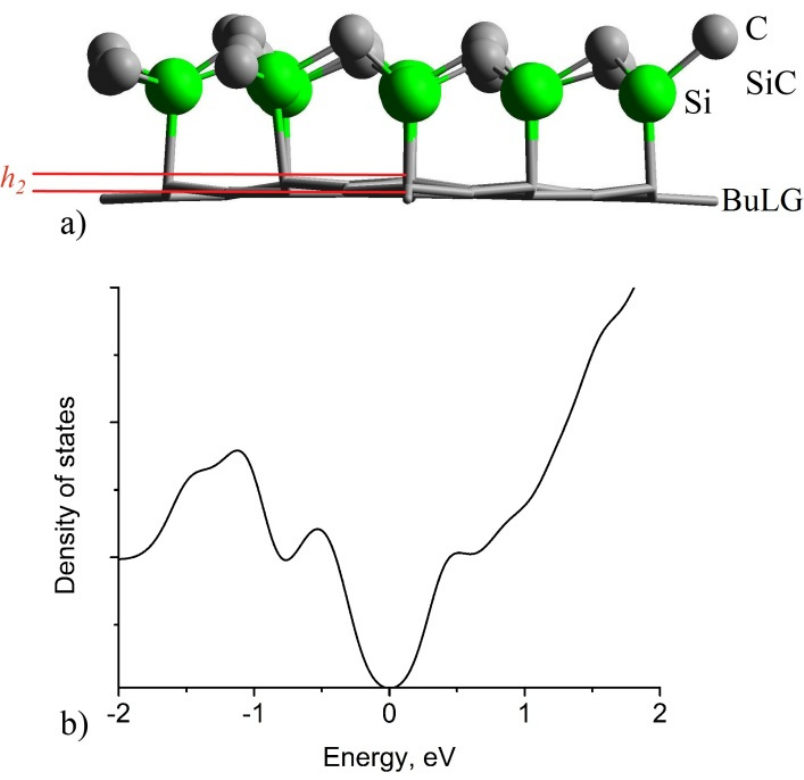}
}
% If not, use
%\vspace{5cm}       % Give the correct figure height in cm
\caption{ a) The structure SiC/BuLG viewed along the direction perpendicular of the bisector of the angle between {\bf A} and {\bf B} axis, $h_2$ = 17.5 ± 2.5 pm; b) The dispersion of the density of states.    }
\label{fig4}       % Give a unique label
\end{figure}

\begin{figure}
% Use the relevant command for your figure-insertion program
% to insert the figure file.
% For example, with the option graphics use
\resizebox{0.4\textwidth}{!}{%
  \includegraphics{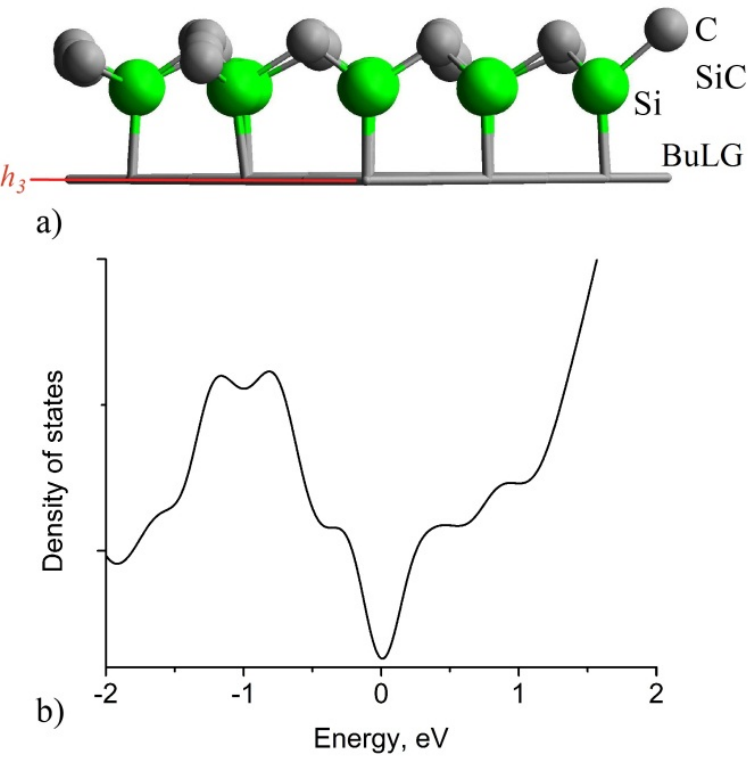}
}
% If not, use
%\vspace{5cm}       % Give the correct figure height in cm
\caption{ a) The structure SiC/ non-corrugated BuLG viewed along the direction perpendicular to the bisector of the angle between {\bf A} and {\bf B} axis, and height of  $h_3$ = 0; b) The dispersion of the density of states.     }
\label{fig5}       % Give a unique label
\end{figure}

The radial distribution function of atoms from BuLG and SiC are presented in Fig. \ref{fig2}c and d. The first maximum at 143 pm corresponds to bonds between sp$^2$ carbon atoms as it is for a single graphene sheet - see Fig. \ref{fig2}c. The second maximum at 154 pm corresponds to single bonds between sp$^2$ and sp$^3$ carbon atoms. For a planar graphene sheet, only a single maximum at 142 pm is observed, Fig. \ref{fig1}a. The mean distance between the silicon atoms and the sp$^3$ carbon atoms, covalently bonded to them is 196 pm - Fig. \ref{fig2}d. The corrugation period of the BuLG is in the interval of 250 - 260 pm. This value is close to the distance between the aligned silicon atoms of 6H-SiC (0001) surface, which is 267 pm (the height of an equilateral triangle with vertices three nearest silicon atoms). The corrugation of the BuLG is governed by the repeating sp$^3$ carbon atoms, covalently bonded to silicon ones. The height of the corrugation of BuLG ($h_1$) is in the range of 35.0 ± 5.0 pm. The energy gap (difference between HOMO and LUMO maxima) is $\Delta$E = 1.14 eV. Systems with different degrees of corrugation are simulated in order to study the influence of the corrugation on the energy gap. A system with $h_2$=17.5 ± 2.5 pm denoted as semi-corrugated and presented in Fig. \ref{fig4}a, has an energy gap of 0.74 eV. The density of states dispersion for the semi-corrugated system is presented in Fig. \ref{fig4}b. A non-corrugated system - when the buffer graphene sheet remains flat, is also studied (Fig. \ref{fig5}a) and is found that its energy gap (HOMO-LUMO difference) is $\Delta$E = 0.29 eV. The density of states is presented in Fig. \ref{fig5}b. The results of the simulations (interatomic distances, the corrugation height and the energy gaps) are summarized in Table \ref{table2}.

\begin{figure}
% Use the relevant command for your figure-insertion program
% to insert the figure file.
% For example, with the option graphics use
\resizebox{0.4\textwidth}{!}{%
  \includegraphics{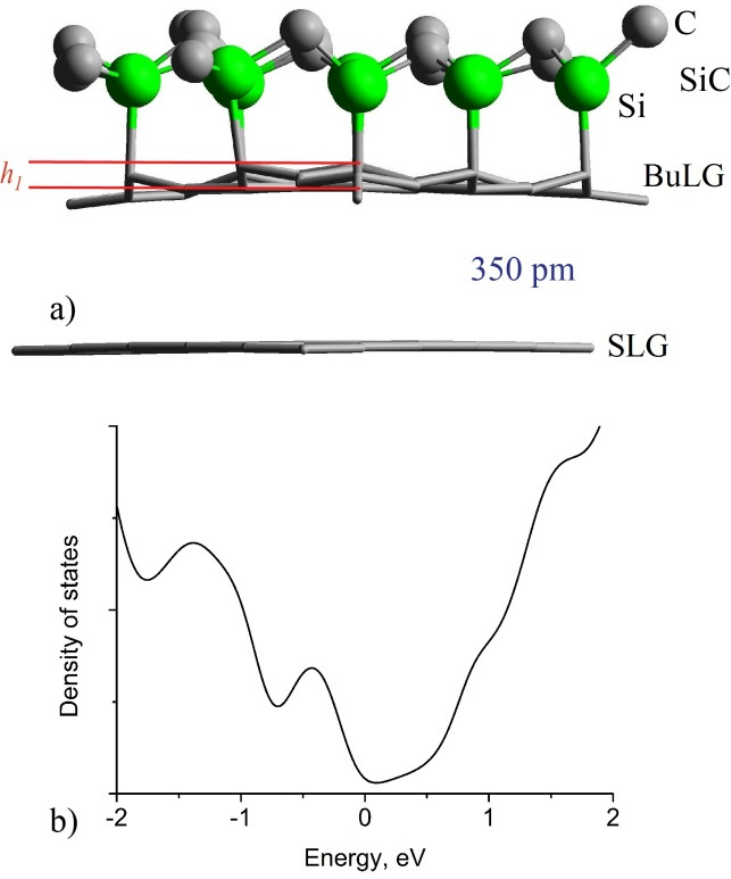}
}
% If not, use
%\vspace{5cm}       % Give the correct figure height in cm
\caption{ a) The structure SiC/ bilayer graphene viewed along the direction perpendicular to the bisector of the angle between {\bf A} and {\bf B} axis. SLG denotes the second graphene layer; b) The dispersion of the density of states.     }
\label{fig6}       % Give a unique label
\end{figure}

A second graphene layer was added to the system, in order to study its interaction with the corrugated BuLG - Fig. \ref{fig6}a. The geometry optimization yielded AB stacked graphene layers with distance between BuLG and the second layer graphene (SLG) of 350 pm. RDF of the second layer is presented in Fig. \ref{fig6}b with first maxima at 145 pm. Тhe second layer is not completely planar and a small corrugation is found with height of about 2 ± 0.5 pm. The energy gap in the bilayer graphene (a corrugated BuLG and a SLG), bonded to SiC, is $\Delta$E = 0.19 eV, Fig. \ref{fig6}c. When a third graphene layer is added to the system, the energy gap is reduced additionally to $\Delta$E = 0.09 eV.

\begin{table*}[b]
\centering
\begin{tabular}{|c|c|c|c|c|}
\hline
  &Corr. 
BuLG in C/SiC	& Semi-corr. 
BuLG in C/SiC &	Non-corr. 
BuLG &	2nd layer \\
\hline
C-C (sp$^2$) in BuLG in pm &	143	&144&	144&	145\\
\hline
C-C (sp$^3$) in BuLG in pm&	154&	153&	153& 	- \\
\hline
Si-C (sp$^3$ C) in pm&	196&	196&	196&	-\\
\hline
Corrugation height in pm&	35.0 ± 5.0&	17.5 ± 2.5&	0.0	&2 ± 0.5\\
\hline
Energy gap ($\Delta$E) in eV	& 1.14&	0.74&	0.29&	0.19\\
\hline
\end{tabular}
\caption{Summary of the data obtained by numeric simulations of the two systems: graphene bonded to SiC (structural models shown in Figures. \ref{fig2}a, \ref{fig2}b, \ref{fig4}a, \ref{fig5}a and \ref{fig6}a).}
\label{table2}
\end{table*}

\section{Conclusions}

We studied the influence of corrugations on the energy gap in graphene as a possible route for making graphene (a zero band gap material) suitable for microelectronics and especially in the construction of field effect transistors for logical applications, where a semiconducting material for the channel is a necessity (ON/OFF state).
 
With the help of {\it ab initio} simulations, we establish that the energy gap of a graphene sheet is directly proportional to the height of corrugation and inversely proportional to its period. The covalent interactions between the graphene and a SiC substrate cause a corrugation of the buffer layer (BuLG) with a height of $h$ = 35.0 ± 5.0 pm and open an energy gap $\Delta$E = 1.14 eV (corrugation of 17.5 ± 2.5 pm has a band gap of $\Delta$E = 0.74 eV; while the non-corrugated system has a band gap of $\Delta$E = 0.29 eV due to the electronic effect from the SiC substrate). The numerically obtained band gap is consistent with the experimental results $\Delta$E $>$ 0.5 eV \cite{[8]} for a similar system. It should be noted that the energy gap opens as a result of the synergy between two effects: i.) the first one being the corrugation caused by the carbon atoms covalently bonded to the aligned silicon atoms of 6H-SiC (0001) surface, with a period of 250 - 260 pm; ii.) the second one being the electronic effects caused by the change of hybridization of these covalently bonded carbon atoms (from sp$^2$ to sp$^3$) and the corresponding removal of p electron density from the delocalized system. 

As an example of non-covalent interaction, we convey the binding of a second graphene layer ontop of the buffer one. In this case, the electronic structure of the system is influenced by the $\pi - \pi$ interaction between the layers in addition to the corrugation of buffer layer. The numerical simulations show that the second layer is not completely planar, but has a small corrugation with a height of 2 ± 0.5 pm. The C-C interatomic distance is 145 pm, 3 pm longer than that of graphene. The energy gap of the two layer system is relatively small and equal to 0.19 eV. It is worth noting that the simulations of a system with three graphene layers point to a further decrease in the energy gap to $\Delta$E = 0.09 eV, like the one of AB - stacked bi-layer graphene after geometry optimization, $\Delta$E = 0.07 eV.

\section{Acknowledgements}
The authors gratefully acknowledge financial support by the National Science Fund of
Bulgaria under grant DN18/9-11.12.2017.

\section{Authors contributions}
S.K. performed the CP2K simulations. V.A. developed the concept of modulation of the forbidden gap of graphene by a corrugation and was responsible for sections 3. Motivation and 4. Main hypothesis. H.A. performed the VASP simulations. T.M. suggested the physical models, how to prove the suitability of the main hypothesis and was responsible for building the structure model/s and its compatibility. S.K. wrote the manuscript with input from all authors.
%
% BibTeX users please use
% \bibliographystyle{}
% \bibliography{}
%
% Non-BibTeX users please use

\end{document}